\documentstyle[11pt,preprint]{aastex}

\shorttitle{Cepheid in a LMC eclipsing binary system}
\shortauthors{Cassisi \& Salaris}

\begin{document}

\title{A Classical Cepheid in a LMC eclipsing binary: evidence of shortcomings in current stellar evolutionary models? }

\author{S. Cassisi\altaffilmark{1} and 
M. Salaris\altaffilmark{2}
}

\altaffiltext{1}{INAF-Osservatorio Astronomico di Collurania,  Via M. Maggini, s.n., 64100 Teramo, Italy; cassisi@oa-teramo.inaf.it}
\altaffiltext{2}{Astrophysics Research Institute, Liverpool John Moores University, Twelve Quays House, Birkenhead CH41 1LD, 
UK; ms@astro.livjm.ac.uk}

\begin{abstract}
The recent discovery and analysis of a classical Cepheid in the well detached, double-lined, 
eclipsing binary OGLE-LMC-CEP0227, has provided the first determination of the 
dynamical mass of a classical Cepheid variable to an unprecedented 1\% accuracy. We show here that modern 
stellar evolution models widely employed to study Galactic and extragalactic stellar systems,   
are able to match simultaneously mass and radius (and effective temperature) 
of the two components with a single value for the age of the system, without 
any specific fine-tuning,  assuming the typical metallicity of LMC Cepheids. 
Our conclusion is that there is no  obvious discrepancy between dynamical 
and evolutionary masses for the Cepheid star in this system, contrary  
to previous claims of an overestimate of the Cepheid mass by stellar evolution theory. 
 \end{abstract}

\keywords{binaries: eclipsing --- stars: evolution --- stars: fundamental parameters --- stars: variables: Cepheids}

\section{Introduction} \label{sec:intro}

Classical Cepheid variables have been the subject of several theoretical and observational studies, due to their 
fundamental role in establishing the extragalactic distance scale \cite[see, e.g.][and reference therein]{fr01, s06}. 
The empirical determination of their pulsational properties provides also sets of 
observational benchmarks for testing both pulsational and stellar evolution models.
One well known problem related to the theory of classical Cepheids is 
the discrepancy between mass estimates obtained from pulsational and stellar evolution models. 
Specifically, by comparing results of theoretical pulsation calculations with the period, 
the estimated mean absolute magnitude and effective temperature of a Cepheid variable, a \lq{pulsational mass}\rq\ 
can be derived. At the same time, a comparison of stellar evolution models with just the 
estimated mean absolute magnitude and effective temperature, a so-called  \lq{evolutionary mass}\rq\ 
can be assigned to the same object. 
Early investigations by \citet{stobie}, \citet{cogan} and \citet{rodgers} discovered that these  
two independent mass estimates did not agree, the pulsational masses being smaller by 20-40\%. 
This disagreement has been named the \lq{Cepheid Mass Discrepancy}\rq. 

The first fundamental step towards solving this problem was discussed by \citet{andreasen}, who suggested that an increase of 
a factor of $\sim2.5$ of the radiative opacities in the temperature regime $1.5<(T/10^5~K)<8$ would be able to remove almost completely the 
discrepancy. This increase was nicely confirmed years later 
by accurate opacity computations performed by the OPAL group \citep{rogers92}. 
Following the implementation of these new opacities in stellar model computations, there has been a large number of 
studies devoted to re-investigate the mass discrepancy problem \citep[see, e.g.][and references therein]
{keller02,keller06,caputo05,evans07}. These analyses concluded that the discrepancy was significantly 
reduced to about 10-15\% in mass. 

Attempts to further reduce the disagreement from the point of view of stellar evolution modelling have invoked the inclusion of 
mass loss, and an increase of the convective core mass during 
the core H-burning stage \citep[see, e.g.,][and references therein]{caputo05,bono06,keller} compared to the values 
determined from the Schwarzschild criterion. 
\citet{keller}, and more recently 
\citet{hilding10}, have shown that mass loss does not appear a viable solution because the required mass loss efficiency 
would not only be extremely large, but also its dependence on the stellar mass would be at odds with independent empirical constraints. 
\citet{hilding10} have also presented compelling evidence to support the idea that a moderate increase 
of the convective core size during the main sequence would be able to eliminate completely the mass discrepancy. 
The proposed increase of the convective core mass may be 
achieved as a consequence of a change in the adopted physical inputs,  
and/or the inclusion of additional physical processes like rotational mixing and mild convective core overshooting. 
Major modifications to the stellar model input physics do not seem realistic \citep[see, e.g.][]{cassisi}, whereas 
a host of observational data -- independent of Cepheid stars -- 
on, e.g., eclipsing binary systems with at least one component on the main sequence or near the turn off, 
the turn off morphology in  
color-magnitude-diagrams of open clusters, star counts along the main sequence and central He-burning sequences in star clusters 
\citep[see, e.g.][and references therein]{andersen, pietrinferni04}, 
demonstrate the need to extend the main sequence convective cores 
beyond the formal Schwarzschild boundary. The added extension of the convective core size, $\Lambda_{cco}$,  
is often parametrized in terms of $\Lambda_{cco}=\lambda{H_P}$, where $H_P$ is the 
pressure scale height at the Schwarzschild boundary and $\lambda$ is a free parameter. 
The current, more widely used libraries of stellar models all include an extension of 
the convective core size beyond the Schwarzschild boundary
\citep[see, e.g.,][and references therein]{pietrinferni04, bertelli, dotter}.

\citet{pietrzynski} have very recently provided
the first accurate determination of the 
dynamical mass of a classical Cepheid variable in a well detached, double-lined, 
eclipsing binary in the Large Magellanic Cloud (OGLE-LMC-CEP0227). 
Thanks to the availability of a high-quality photometric dataset, the spectroscopic follow-up, and the 
near-perfect characteristics of the system to derive accurate masses for both components, \citet{pietrzynski} 
were able to estimate the mass of 
the pulsator to an unprecedented 1\% precision.
The discovery and analysis of this binary system is a pivotal step forward in testing the accuracy of both pulsation and 
stellar evolution theory. 
\citet{pietrzynski} emphasize that their dynamical estimate of the Cepheid mass 
is in good agreement with the estimated pulsational mass, whilst they suggest the existence of a discrepancy with the mass predicted 
by stellar evolution models. In view of the high precision of the parameters of this binary system, and the huge effort devoted in this 
last decade towards improving stellar evolution models, we consider that the 
estimate of the Cepheid evolutionary mass for this system deserves a more specific investigation. 
The accurate determination of masses and radii for both components allows us to test 
for the first time in a more direct 
way the accuracy of theoretical evolutionary masses of Cepheid stars. Following a standard and powerful 
methodology of eclipsing-binary studies \citep[i.e.,][]{andersen}, we will test whether in the 
age-radius diagram, evolutionary models calculated for the empirical values of the mass of the two components and a 
typical LMC composition, can match the observed radii for a common value of the age.
With this procedure we want to address the following question: 
is it possible to reproduce the mass/radius constraints of both components of OGLE-LMC-CEP0227 by employing a set of widely used 
off-the-shelf modern stellar evolution models, without any fine-tuning for this specific object?

The next section describe briefly the models employed in our analysis and presents a comparison with 
the results for OGLE-LMC-CEP0227. A final discussion closes the paper.

\section{The comparison between stellar evolution theory and observations} \label{sec:evo}

We have employed the same state-of-the-art input physics and stellar evolution code 
adopted for building the widely used BaSTI stellar model archive\footnote{The BaSTI stellar evolutionary model archive is available at the following URL address: http://www.oa-teramo.inaf.it/BASTI.} \citep[see][for details]{pietrinferni04}, to 
compute a series of models with masses between $\sim$4.1 and $\sim$4.2~$M_{\odot}$, appropriate for the  
OGLE-LMC-CEP0227 components. 
The BaSTI archive 
has been extensively tested against observations of local stellar populations and eclipsing binary 
systems \citep[see, e.g.][and references therein]{tomasella}, and is widely employed 
in studies of Galactic and extragalactic resolved star clusters \citep[see, e.g.][for just a few examples]{deangeli, mbn, amarin},  
in the determination of Star formation history and chemical enrichment history of 
resolved Local Group galaxies \citep[see, e.g.,][]{carrera}, and also in studies of integrated properties 
of extragalactic stellar populations \citep[see, e.g.,][]{carter}.

We adopted in our calculations 
the typical initial chemical composition of LMC stars, Z=0.008 and Y=0.256 
-- that corresponds to [Fe/H]=$-$0.35 for the solar (Z/X) ratio by \citet{gn93} -- e.g. the same 
composition used by \citet{pietrzynski} to determine the pulsational mass of the Cepheid variable.  
The convective cores along the main sequence have been extended by an amount $\Lambda_{cco}=0.2{H_P}$, as in 
the BaSTI models for this mass range. 
As discussed in \citep{pietrinferni04} the calibration of $\Lambda_{cco}$ in 
the BaSTI models is based on the morphology of color-magnitude-diagrams of open clusters, and is completely independent of 
the evolution in the Cepheid instability strip. Finally, all models 
presented in this work have been computed by adopting a \citet{reimers} mass-loss law, with the free parameter 
$\eta$ set to 0.4, and a solar mixing length calibration, as in the BaSTI database \citep{pietrinferni04}.

Our analysis detailed below shows that the BaSTI stellar 
evolutionary models widely employed to study Galactic and extragalactic stellar populations 
are able to match the mass and radius of  
both components of the binary system studied by \citet{pietrzynski} with a single value for the age of the system, 
without any specific fine-tuning.   
This implies that there is no significant discrepancy between dynamical and evolutionary masses for the Cepheid star 
in this system. This is a result opposite to the conclusions by 
\citet{pietrzynski}, who suggest an overestimate of Cepheid masses by stellar evolution theory.

The relevant parameters of the eclipsing binary system derived by \citet{pietrzynski} are reported in table~\ref{Ecltab}. 
The values  of the stellar radius and $T_{eff}$ for the Cepheid \citep[the primary component A, as labelled in][]{pietrzynski} 
are mean values suitable to be compared with results from stellar models in hydrostatic equilibrium.

\begin{figure}
%\epsscale{1.0, rotate=90}
%\plotone{eclip_cef_rtime.eps}
\centering
\includegraphics[width=12.3truecm, height=12.3truecm]{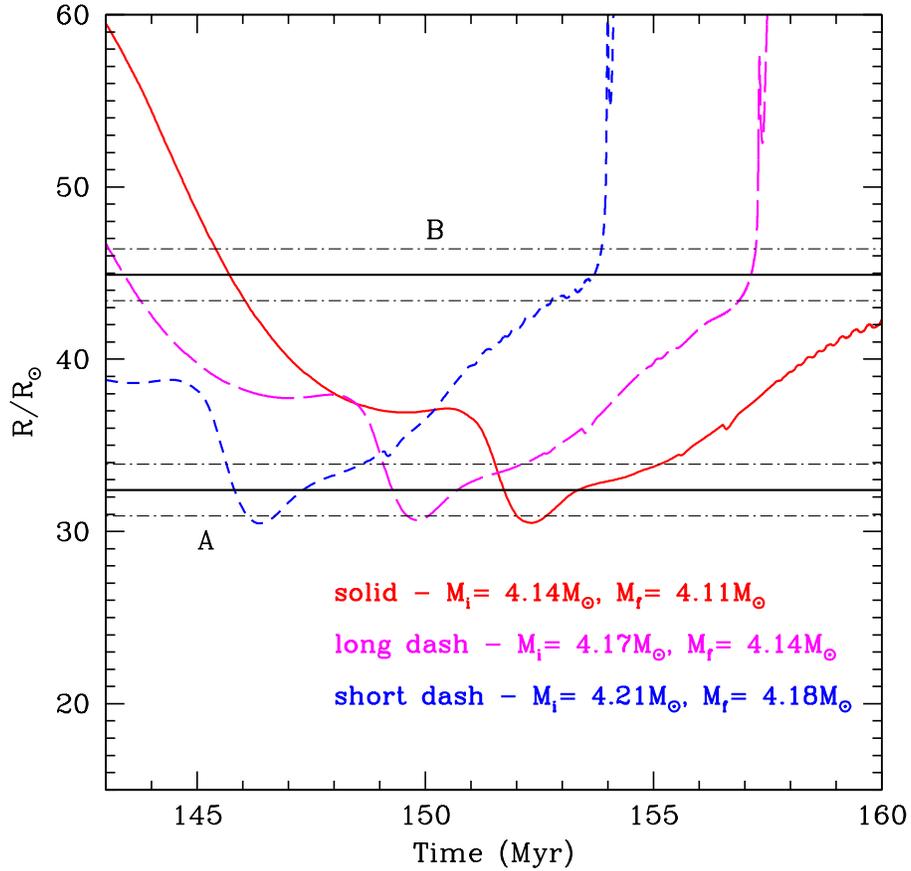}
\caption{The age - radius diagram: empirical values for the radii of the 
primary (labelled as A) and secondary (labelled as B) components are displayed as solid horizontal lines.
The horizontal dash-dotted lines mark the boundaries of the 1$\sigma$ error regions.
Three theoretical models with different values of the initial main sequence mass ($M_i$) and corresponding mass 
during the He-burning phase ($M_f$) are also displayed. The mass loss is efficient mainly along the red giant 
phase before central He-ignition, and reduces the stellar mass by a constant amount, equal to 0.03$M_{\odot}$.
The three selected values of $M_i$ and $M_f$ are all within the 1$\sigma$ error bar of the empirical estimates 
(see Table~\ref{Ecltab}).
\label{fig:rad}}
\end{figure}

Figure~\ref{fig:rad} displays a comparison theory-observations in the age-radius diagram. This 
diagram ensures the most direct test between theoretical and empirical quantities, without the need to apply  
color - $T_{eff}$ relations and bolometric corrections to the output of stellar evolution calculations. 

A satisfactory fit with stellar evolution models of the appropriate mass 
must reproduce -- within the error bars -- the radii of the two components for a common age. 
The three selected values for the intial mass of the models ($M_i$) and 
the corresponding mass during the central He-burning phase ($M_f$) are all within the 1$\sigma$ error bar of the 
empirical values. Mass loss implemented with the Reimers formula is efficient mainly along the red giant 
phase before central He-ignition, and reduces the stellar mass by a constant amount, equal to $\sim$0.03$M_{\odot}$. 
Analogous computations 
with $\eta$=0 (i.e. with inefficient mass loss)  provide essentially the same results 
in the $R - age$ diagram, the only difference being that in this case $M_i=M_f$ by definition.
 
Even a cursory analysis of Fig.~\ref{fig:rad} shows that stellar evolution models provide masses in complete agreement with the 
results from the eclipsing binary analysis. Considering the model with  
$M_i$=4.14 $M_{\odot}$  ($M_f$=4.11 $M_{\odot}$) for the Cepheid component, and the one with $M_i$=4.21 $M_{\odot}$ 
($M_f$=4.18 $M_{\odot}$) for the companion, the observed radii are 
reproduced within the 1$\sigma$ error bars for a common age of $\sim$154~Myr. 
These results demonstrate that there is no discrepancy between evolutionary and 
dynamical mass of the Cepheid star in this binary system, when stellar evolution models include updated input physics and 
the extension of the convective cores satisfy independent constraints from open cluster color-magnitude-diagrams.

\begin{figure}
%\epsscale{1.0, rotate=90}
%\plotone{{eclip_cef_mass.eps}
\centering
\includegraphics[width=12.3truecm, height=12.3truecm]{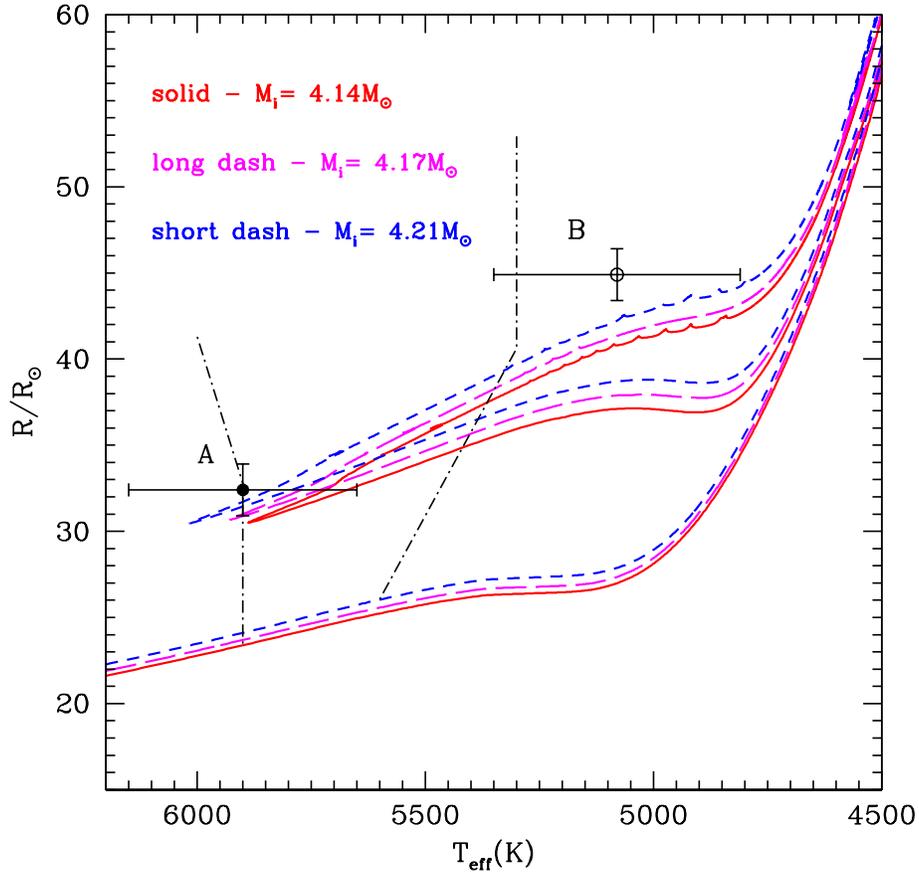}
\caption{The radius - effective temperature diagram: the stellar evolution models are the same as in Fig.\ref{fig:rad}, 
while the filled circles and associated error bars correspond to the empirical estimates for the binary system components. 
The location of both the Fundamental Blue Edge and Fundamental Red Edge of the Cepheid instability strip are also plotted.
\label{fig:temp}}
\end{figure}

Figure~\ref{fig:temp} displays an additional comparison in the $R - T_{eff}$ diagram. 
The stellar models displayed are the same as in Fig.~\ref{fig:rad} and the agreement with observations is again 
excellent for the same combinations of $M_i$ values that satisfies the $R - age$ diagram. 
We have plotted also the Fundamental Blue edge (FBE) and Fundamental Red edge (FRE) of the Cepheid instability strip as 
provided by \citet{fiorentino} pulsational models, for a metallicity of Z=0.008, and a mixing length equal to 1.5$H_p$, 
their reference choice for the treatment of the superadiabatic layers.
The two components appear to have been captured - as also remarked by \citet{pietrzynski} - in a very 
short-lived evolutionary stage, i.e. when they move along the brighter branch of the blue loop 
during the core He-burning stage. By examining Fig.~\ref{fig:temp} we can add that the A component is located at the boundary 
between the fundamental and the first overtone region within the instability strip, whereas the B component appears to have left 
the instability strip since a short time. The fact that the B component is not a variable star, and its location 
with respect to the FRE, poses an important constraint on the efficiency of superadiabatic 
convection in the pulsational models: any substantial decrease of the mixing length adopted in the pulsational models would shift the 
FRE towards lower $T_{eff}$ \citep[e.g.][]{fiorentino}, moving the B component 
within the instability strip, at odds with the observations. 

Before closing this section we wish to 
comment briefly about the effect of our assumptions about metallicity and mass-loss, for there are 
no spectroscopic estimates of the chemical composition of OGLE-LMC-CEP0227 components, and no strong constraints 
on the mass-loss rate of Cepheids and their progenitors. 
Our analysis, as well as the value determined by \citet{pietrzynski} for the pulsational mass of 
the Cepheid component, is based on the assumption of a 'typical' iron abundance [Fe/H]=$-$0.35, that is 
consistent with the mean value determined spectroscopically by \citet{romaniello} for a sample 
of 22 LMC Cepheids. These authors also find a spread of [Fe/H] values for their sample, with 
[Fe/H] ranging between $-$0.62 and $-$0.10~dex. Given the lack of [Fe/H] estimates for 
OGLE-LMC-CEP0227 components, we have tested to what extent assumptions on their metallicity are critical 
for the outcome of this analysis. We found that varying [Fe/H] by $\pm$0.10~dex (i.e. varying  
Z between 0.006 and 0.01, and Y using the He-enrichment ratio 
$\Delta$Y/$\Delta$Z=1.4, as in the BaSTI archive) around the reference value 
still ensures consistency between theory and observations in the age-radius diagram. 
Values of [Fe/H] outside this range prevent models -- with mass consistent with the empirical values -- 
from crossing the Cepheid instability strip during the 
central He-burning phase. As a consequence, one would need in this case to fine-tune the evolutionary calculations to match 
this system, for example by including overshooting from the bottom of the convective envelopes, that favours the 
development of loops in the Hertzsprung-Russell diagram during the central He-burning phase.

As for the mass-loss assumptions, we have already remarked that neglecting mass loss or employing  
the Reimers prescription with $\eta$=0.4 (a standard choice for the value of this free parameter) do provide 
consistency between evolutionary and empirical masses. Observations have not determined yet 
conclusive and stringent bounds on mass-loss rates for Cepheids and their progenitors. 
Very recently, \citet{marengo} have found some indications that the Cepheid prototype $\delta$ Cephei may be 
currently losing mass, at a rate in the range of $\approx 5 \ 10^{-9}$ to $6 \ 10^{-8} M_{\odot} \ yr^{–1}$.
As an extreme test, we have calculated models employing the upper limit of this mass-loss rate range during the 
central He-burning phase, and our standard assumptions on 
the binary metallicity. The models lose $\approx 2M_{\odot}$ during the 
He-burning phase, and the mass loss efficiency is so large that the stellar models
show a very anomalous path in the H-R digram.
Such a high mass-loss rate -- if conclusively established and, moreover, if typical of Cepheids also in the LMC -- 
would be compatible with standard stellar evolution models only in case of episodic mass-loss events, 
that do not change the total mass of the star by more than a few hundredths of solar masses.

\section{Conclusions} \label{sec:close}

The recent study by \citet{pietrzynski} of a classical Cepheid in the well detached, double-lined, 
eclipsing binary OGLE-LMC-CEP0227, has provided the first very accurate determination of the 
dynamical mass of a classical Cepheid variable, and its radius.
Our analysis has shown that stellar evolutionary models widely employed to study Galactic and extragalactic stellar populations 
are able to match -- assuming the typical metallicity of LMC Cepheids --  the derived mass and radius of both components 
with a single value for the age of the system, without any specific fine-tuning.    
Our result implies that there is no discrepancy between dynamical and evolutionary masses for the Cepheid star 
in this system, when current stellar models - calculated with up-to-date stellar physics 
and extensively tested against independent empirical benchmarks - are used in conjunction with the method 
employed in this analysis.

It is also worth noticing that the agreement between evolutionary masses and high precision  
dynamical masses of Cepheids discussed in this paper, is based on just a single eclipsing binary system. The 
discovery and analysis of more systems similar to OGLE-LMC-CEP0227 -- in conjunction with 
direct measurements of their chemical composition -- will be enormously beneficial to the fields of 
stellar evolution and pulsation theory.

\acknowledgments

This research has made use of NASA's  
Astrophysics Data System Bibliographic Services, which is operated by the Jet
Propulsion Laboratory, California Institute of Technology, under contract with the
National Aeronautics and Space Administration. S.C. acknowledges the financial
support of PRIN-INAF 2010 (PI: R. Gratton),
ASI grant ASI-INAF I/016/07/0, and the Italian Theoretical Virtual Observatory
Project (PI: F. Pasian).

\newpage

%%%%%%%%%%%%%%%%%%%%%%%%%%%%%%%%%%%%%%%%%%%%%%%%%%%%%%%%%%%%%%

\begin{table}[ht!]
%\begin{table}
\centering
\caption{Parameters of OGLE-LMC-CEP0227}
\begin{tabular}{lcc} 
\hline
 & Primary (A)&  Secondary (B)\\  
\hline\hline
Mass (${\rm M/M_{\odot}}$) &  4.14$\pm$0.05 & 4.14$\pm$0.07 \\
Radius (${\rm R/R_{\odot}}$)  & 32.4$\pm$1.5  & 44.9$\pm$1.5 \\ 
${\rm T_{eff} (K)}$ &  5900$\pm$250 &  5080$\pm$270\\ 
\hline 
\hline
\label{Ecltab}
\end{tabular}
\end{table}

\newpage

%%%%%%%%%%%%%%%%%%%%%%%%%%%%%%%%%%%%%%%%%%%%%%%%%%%%%%%%%%%%%%
%%%%%%%%%%%%%%%%%%%%%%%%%%%% FIG 1  %%%%%%%%%%%%%%%%%%%%%%%%%%%%%%

%%%%%%%%%%%%%%%%%%%%%%%%%%%%%%%%%%%%%%%%%%%%%%%%%%%%%%%%%%%%%%

\newpage
%%%%%%%%%%%%%%%%%%%%%%%%%%%%%%%%%%%%%%%%%%%%%%%%%%%%%%%%%%%%%%
%%%%%%%%%%%%%%%%%%%%%%%%%%%% FIG 2  %%%%%%%%%%%%%%%%%%%%%%%%%%%%%%

%%%%%%%%%%%%%%%%%%%%%%%%%%%%%%%%%%%%%%%%%%%%%%%%%%%%%%%%%%%%%%

\end{document}